\documentclass[
reprint,
superscriptaddress,
showpacs,
amsmath,amssymb,
aps,
prb,
floatfix
]{revtex4-1}

\usepackage{graphicx}
\usepackage{dcolumn}
\usepackage{bm}

\begin{document}
\preprint{APS/123-QED}
\title{Devil's staircases in the IV characteristics of superconductor/ferromagnet/superconductor Josephson junctions}
\author{M. Nashaat}
\affiliation{Department of Physics, Cairo University, Cairo, 12613, Egypt}
\affiliation{BLTP, JINR, Dubna, Moscow Region, 141980, Russia}
\author{A. E. Botha}
\affiliation{Department of Physics, University of South Africa, Science Campus, Private Bag X6, Florida Park 1710, South Africa}
\author{Yu. M. Shukrinov}
\email{shukrinv@theor.jinr.ru}
\affiliation{BLTP, JINR, Dubna, Moscow Region, 141982, Russia}
\affiliation{Department of Physics, University of South Africa, Science Campus, Private Bag X6, Florida Park 1710, South Africa}
\affiliation{Dubna State University, Dubna, Russian Federation}

\date{\today}

\begin{abstract}
We study the effect of coupling between the superconducting current and magnetization in the superconductor/ferromagnet/superconductor Josephson junction under an applied circularly polarized magnetic field.  Manifestation of ferromagnetic resonance in the frequency dependence of the amplitude of the magnetization and the average critical current density is demonstrated. The IV characteristics  show subharmonic steps that form devil's staircases, following a continued fraction algorithm. The origin of the found steps is related to the effect of the magnetization dynamics on the phase difference in the Josephson junction. The dynamics of our system is described by a generalized RCSJ model coupled to the Landau-Lifshitz-Gilbert equation. We justify analytically  the appearance of the fractional steps in IV characteristics of the superconductor/ferromagnet/superconductor Josephson junction.
\end{abstract}
\maketitle

\section{Introduction}
An important challenge, in superconducting spintronics dealing with the Josephson junctions coupled to magnetic systems, is the  achievement of electric control over the magnetic properties by the Josephson current and its counterpart, i.e. the achievement of magnetic control over the Josephson current.\cite{vzutic2004spintronics,linder2015superconducting,golubov2004current,mai2011interaction} In some systems, spin-orbit coupling plays a major role in the attainment of such control.\cite{buzdin2008direct} For example, a recent study showed a full magnetization reversal in a superconductor/ferromagnet/superconductor (S/F/S) structure, with spin-orbit coupling, by adding an electric current pulse.\cite{shukrinov2017magnetization}  Such a reversal may be important for certain applications.~\cite{shukrinov2017magnetization}  Another approach was followed in Refs.~\onlinecite{cai2010interaction,cai2013reversal}, where the authors demonstrated the interaction of a nanomagnet with a weak superconducting link and the reversal of single domain magnetic particle magnetization by an ac field. The superconducting current of a Josephson junction (JJ) coupled to an external nanomagnet driven by a time-dependent magnetic field both without and in the presence of an external ac drive were studied in Ref.~\onlinecite{ghosh}. The authors showed  the existence of Shapiro-type steps in the IV characteristics of the JJ subjected to a voltage bias for a constant or periodically varying magnetic field and explored the effect of rotation of the magnetic field and the presence of an external ac drive on these steps. Furthermore, a uniform  precession mode (spin wave) could be excited by a microwave magnetic field, at ferromagnetic resonance (FMR), when all the elementary spins precess perfectly in phase.\cite{ounadjela2003spin} Finally, coupling between the Josephson phase and a spin wave was studied in the series of papers.\cite{weides2006,pfeiffer2008static,hikino2011ferromagnetic,wild2010josephson,kemmler2010magnetic,volkov2009hybridization,mai2011interaction}

In Josephson junctions driven by external microwave radiation the Shapiro steps~\cite{shapiro} that appear in the IV characteristics can form the so-called devil's staircase (DS) structure as a consequence of the interplay between Josephson plasma and applied frequencies.\cite{ben-jacob,shukrinov2013devil,shukrinov2014structured,sokolovic2017devil} The DS structure is a universal phenomenon and  appears in a wide variety of different systems, including infinite spin chains with long-range interactions,\cite{nebendahl13} frustrated quasi-two-dimensional spin-dimer systems in magnetic fields,\cite{takigawa13} and even in the fractional quantum Hall effect.\cite{hriscu13} In Ref.~\onlinecite{tang2018} the authors considered symmetric dual-sided adsorption, in which identical species adsorb to opposite surfaces of a thin suspended membrane, such as graphene. Their calculations predicted a devil's staircase of coverage fractions for this widely studied system.\cite{tang2018}  In Ref.~\onlinecite{chen2017} a series of fractional integer size steps was observed experimentally in the Kondo lattice CeSbSe. In this system the application of a magnetic field resulted in a cascade of magnetically ordered states -- a possible devil's staircase. A devil's staircase was also observed in soft-x-ray scattering measurements made on single crystal SrCo$_6$O$_{11}$, which constitutes a novel spin-valve system.\cite{matsuda2015} An extension of the investigation of this problem on the S/F/S Josephson junction might open horizons in this field.

The problem of coupling between the superconducting current and magnetization in the S/F/S Josephson junction attracts much attention today (see Ref.~\onlinecite{linder2015superconducting} and the references therein). An intriguing opportunity is related to the connection between the staircase structure and current-phase relation.\cite{maiti15} Particularly, the manifestation of the staircase structure in the IV characteristics of S/F/S junctions might provide the corresponding information on current-phase relation and, in this case, serve as a novel method for its determination. The appearance of the DS structure and its connection to the current-phase relation in experimental situations has not yet been investigated in detail. It stresses a need for a theoretical model which would fully describe the dynamics of the S/F/S Josephson junction under external fields, features of Shapiro-like steps and their DS staircase structures. In Ref.~\onlinecite{hikino2011ferromagnetic} the Josephson energy in the expression for the effective field was not considered. Consequently, the IV characteristics of the S/F/S junction at FMR only showed current steps at voltages corresponding to even multiples of the applied frequency. The authors related these steps to the interaction of Cooper pairs with an even number of magnons.\cite{hikino2011ferromagnetic}

In this paper we investigate the effect of coupling between the superconducting current and magnetization in the superconductor/ferromagnet/superconductor Josephson junction under an applied circularly polarized magnetic field. Taking into account the Josephson energy in the effective field, we demonstrate an appearance of odd and fractional Shapiro steps in IV characteristics, in addition to the even steps that were reported in Ref.~\onlinecite{hikino2011ferromagnetic}. We demonstrate the appearance of devil's staircase structures and show that voltages corresponding to the subharmonic steps under applied circularly polarized magnetic field follow the continued fraction algorithm.\cite{shukrinov2013devil,shukrinov2014structured,sokolovic2017devil}  An analytical consideration of the linearized model, based on a generalized RCSJ model and Landau-Lifshitz-Gilbert (LLG) equation, including the Josephson energy in the effective field, justifies the appearance of the fractional steps in IV characteristics, in agreement with our numerical results.  We also show the manifestation of ferromagnetic resonance in the frequency dependence of the amplitude of the magnetization and the average critical current density. An estimation of the model parameters shows that there is a possibility for the experimental observation of this phenomenon.

The plan of the rest of the paper is as follows. In Sec. II, we describe the model and present an explicit form of the equations. Ferromagnetic resonance is demonstrated in Sec. III, where the effect of Gilbert damping is shown and  a comparison with the linearized case is presented. This is followed by a discussion of the IV characteristics and observed staircase structures in Sec. IV. In Sec. V we discuss the additional effect of an oscillating electric field on the Shapiro steps. Demonstration of different possibilities of the frequency locking and discussion of the experimental realization of the found effects is presented in Sec. VI. Finally, we conclude in Sec. VII and specify our calculations for linearized case.\cite{supplemental}

\section{Model and Methods}
\begin{figure}[!ht]
	\includegraphics[width=0.9\linewidth]{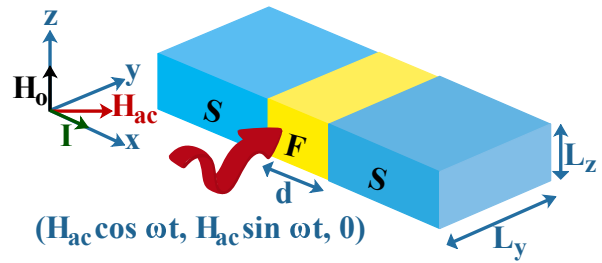}
	\caption{(Color online) Geometry of the S/F/S Josephson junction with cross-sectional area $ L_{y}L_{z}$ in uniform magnetic field $H_0$ and circularly polarized magnetic field $H_{ac}$.}
\label{fig1}
\end{figure}

The geometry of the S/F/S Josephson junction under an applied circularly polarized magnetic field is shown in Fig.~\ref{fig1}. There is a uniform magnetic field of magnitude $H_0$ applied in the $z$-direction. Additionally, a circularly polarized magnetic field, of amplitude $H_{ac}$ and frequency $\omega$, is applied in $xy$-plane. The total applied field is thus $\mathbf{H}(t) = \left( H_{ac}\cos(\omega t), H_{ac}\sin(\omega t),H_0 \right)$.  A bias current $I$ flows in the $x$-direction.
The microwave sustains the precessional motion of the magnetization in the presence of Gilbert damping. The magnetic fluxes in the $z$- and $y$-directions are given by $\Phi_{z}(t)=4 \pi d L_{y} M_{z}(t)/\Phi_{0}$, $\Phi_{y}(t)=4 \pi d L_{z}M_{y}(t)/\Phi_{0}$, where $M_{z}$ and $M_{y}$ are components of magnetization and  $d$ is the thickness of ferromagnet. Using the equation
$\nabla \theta(y,z,t) = - \frac{2 \pi d}{\Phi_{0}} \textbf{B}(t)
\times \textbf{n}$, $\textbf{n}$ is the unit vector in x direction and the fact that two superconductors are thicker than London's penetration depth, we obtain an expression for the gauge-invariant
phase difference, $\theta (y,z,t)=\theta (t)-\frac{ 8 \pi^{2}  d
	M_z(t)}{\Phi _{0}} y+\frac{8 \pi^{2}  d M_y(t)}{\Phi _{0}} z$, where
$\Phi_{0}=h/(2e)$ is the magnetic flux quantum.  Hence, within the framework of the modified RCSJ model, which takes into account the gauge invariance including the magnetization of the ferromagnet,\cite{hikino2011ferromagnetic} the electric current reads
\begin{eqnarray}
I/I^{0}_{c} &=& \frac{ \sin \left(\frac{\pi\Phi_{z}(\tau)}{\Phi
		_{0}}\right)\sin \left(\frac{\pi\Phi_{y}(\tau)}{\Phi
		_{0}}\right)}{(\pi\Phi_{z}(\tau)/\Phi _{0})(\pi\Phi_{y}(\tau)/\Phi
	_{0})}\sin \theta(\tau) \nonumber \\
& &   \hspace*{2.5cm} + \frac{{\rm d}\theta(\tau)}{{\rm d}\tau}
+ \beta_{c} \frac{{\rm d}^{2}\theta(\tau)}{{\rm d}\tau^{2}},
\label{eq1}
\end{eqnarray}
where $\tau=t\omega_{c}$ is the normalized time, $\omega_{c}=2 \pi
I^{0}_{c} R/\Phi_{0}$ is the characteristic frequency, $R$ is the junction resistance, $\beta_{c}=R C \omega_{c}$ is the McCumber parameter, \cite{stewart1968current} and $C$ is the junction capacitance. In the present
paper we will only consider the overdamped case for which $\beta_{c}=0$.

The applied circularly polarized magnetic field in the $xy$-plane causes precession of the magnetization $\textbf{M}$  in the ferromagnetic (FM) layer. The dynamics of the magnetization is described by the LLG equation\cite{ounadjela2003spin}
\begin{equation}
(1+\alpha^{2})\dfrac{{\rm d}\textbf{M}}{{\rm d}t}=-\gamma\textbf{M}\times\textbf{H}_e
- \dfrac{\gamma\alpha}{\lvert\textbf{M}\lvert} \textbf{M}\times
(\textbf{M} \times \textbf{H}_e), \label{eq2}
\end{equation}
where  $\alpha$ is the Gilbert damping, $\gamma$ is the gyromagnetic ratio, and $\mathbf{H}_{e}$ is an effective field. Taking into account that the phase difference depends on the magnetization components, we write the total energy of our system as $E=E_{s} + E_{M} + E_{ac}$, where
\begin{eqnarray}
E_{s\;} &=& -\frac{\Phi_0}{2\pi} \left(\theta(t) - \frac{8 \pi^{2}  d}{\Phi_0}\left( M_z(t) y - M_y(t) z \right) \right) I + \label{eq3} \\
& & E_{J} \left[1-\cos\left( \theta(t) - \frac{8 \pi^{2}  d}{\Phi_0} \left( M_z(t) y - M_y(t) z \right) \right)\right],  \nonumber \\
E_{M} &=& - v H_{0}  M_{z}(t), \nonumber  \\
E_{ac}&=& - v M_{x}(t) H_{ac} \cos (\omega t) - v M_{y}(t) H_{ac} \sin (\omega t). \nonumber
\end{eqnarray}
Here $H_{0}=\omega_{0}/\gamma$, $\omega_{0}$ is the ferromagnetic resonance frequency, and $v$ is the volume. When we switch on $H_0$ and $H_{ac}$, the phase difference starts to depend on $M$, and so does the Josephson energy. The addition of $E_s$ leads to the dependence of the effective field on the ratio $E_J/E_M$, and generalizes the considerations made in Ref.~\onlinecite{hikino2011ferromagnetic}. The effective field is now given by
\begin{equation}
\mathbf{H}_{e}= - \frac{1}{v} \nabla_{\bf{M}} E. \label{eq4}
\end{equation}
In dimensionless form, we write $m=\textbf{M}/M_{0}$,
$M_{0}=|\textbf{M}|$, $\mathbf{h}_{e} = \mathbf{H}_{e}/H_{0}$, $h_{ac}
=H_{ac}/H_{0}$,  $\Omega = \omega /\omega_{c}$, and  $\Omega_{0} =
\omega_{0} /\omega_{c}$,
After integrating the total effective field over the junction area, it has the following form
\begin{eqnarray}
\mathbf{h}_{e}& = & \left( h_{ac} \cos \Omega \tau \right) \hat{\mathbf{e}}_{x} + \left(
h_{ac} \sin \Omega \tau + \varGamma_{yz} \epsilon_{J}\cos \theta
\right) \hat{\mathbf{e}}_{y} \nonumber \\ & & + \left( 1 + \varGamma_{zy}
\epsilon_{J}\cos \theta \right) \hat{\mathbf{e}}_{z},
\label{eq6}
\end{eqnarray}
where $\epsilon_{J} =E_{J}/\left( v M_{0} H_{0} \right)$ and
\begin{eqnarray}
\varGamma_{yz}&=&\frac{ \sin \left(   \phi_{sy} m_{z}\right)}{ m_{y} (
	\phi_{sy} m_{z})} \left[   \cos ( \phi_{sz}
m_{y})-\frac{\sin(\phi_{sz} m_{y})}{(\phi_{sz} m_{y})} \right],\nonumber \\
& & \\ \label{eq7}
\varGamma_{zy}&=&\frac{ \sin \left(   \phi_{sz}
	m_{y}\right)}{ m_{z} ( \phi_{sz} m_{y})} \left[   \cos ( \phi_{sy}
m_{z})-\frac{\sin(\phi_{sy} m_{z})}{(\phi_{sy} m_{z})} \right], \nonumber
\end{eqnarray}
with  $\phi_{sy}$=$4\pi^2 L_{y} d M_{0}/\Phi_{0}$, and
$\phi_{sz}$=$4\pi^2 L_{z} d M_{0}/\Phi_{0}$. If we set $\varGamma_{yz}=\varGamma_{zy}=0$, our system reduces to that of Ref.~\onlinecite{hikino2011ferromagnetic}. We note that the first term for $E_s$ in Eq. (\ref{eq3}), does not contribute to the effective field after integration over the junction area and taking the derivative with respect to the magnetization. The LLG equation in the dimensionless form reads
\begin{equation}
\dfrac{d\mathbf{m}}{d\tau} = - \frac{\Omega_{0}}{(1+\alpha^{2})}\bigg( \mathbf{m} \times
\mathbf{h}_{e} + \alpha \left[ \mathbf{m} \times ( \mathbf{m} \times \mathbf{h}_{e})\right] \bigg)
\label{eq5}.
\end{equation}

The magnetization and phase dynamics of the considered S/F/S Josephson junction is determined by Eqs.~(\ref{eq1}) and (\ref{eq5}). To solve this system and calculate the IV characteristics, we assume a constant bias current and calculate the voltage from the Josephson relation $V(\tau) = {\rm d} \theta /{\rm d} \tau$. We employ a 4th-order Runge-Kutta integration scheme which conserves the magnetization magnitude in time. The dc bias current $I$ is  normalized to the critical current $I^0_c$,  and the voltage $V(t)$ to $\hbar \omega_c/(2e)$. As a result, we find the temporal dependence of the voltage in the JJ at a fixed value of bias current $I$. Then, the current value is increased or decreased by a small amount, $\delta I$ (the bias current step), to calculate the voltage at the next point of the  IV characteristics. We use the final phase and voltage achieved at the previous point of the IV characteristics as the initial condition for the next current point. The average of the voltage $V(\tau)$ is given by $V =\frac{1}{T_{f}-T_{i}}\int^{T_{f}}_{T_{i}}V(\tau) {\rm d}\tau$, where $T_{i}$ and $T_{f}$ determine the interval for the temporal averaging. Further details of the simulation procedure are described in Ref.~\onlinecite{sg-prb11}. The initial conditions for the magnetization components are assumed to be $m_{x}=0$, $m_{y}=0.01$ and $m_{z}=\sqrt{1-m^{2}_{x}-m^{2}_{y}}$, while for the voltage and phase we take zeros. The numerical parameters (if not mentioned) are $\alpha=0.1$, $h_{ac}=1$, $\phi_{sy}=\phi_{sz}=4$, $\epsilon_{J}=0.2$ and $\Omega=\Omega_{0}=0.5$.

\section{Ferromagnetic Resonance}

\begin{figure}[!ht]
	\includegraphics[width=.83\linewidth]{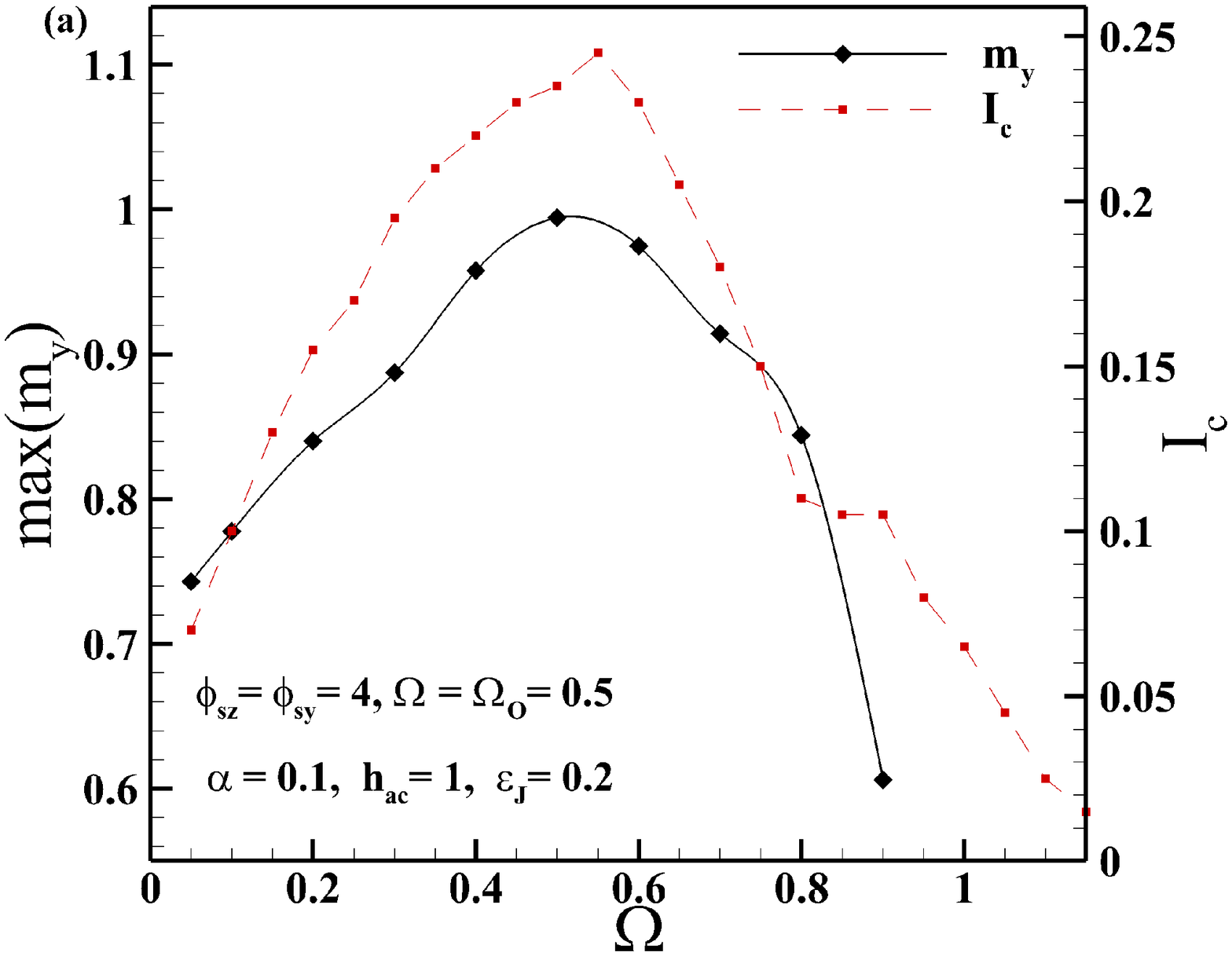}
\includegraphics[width=.83\linewidth]{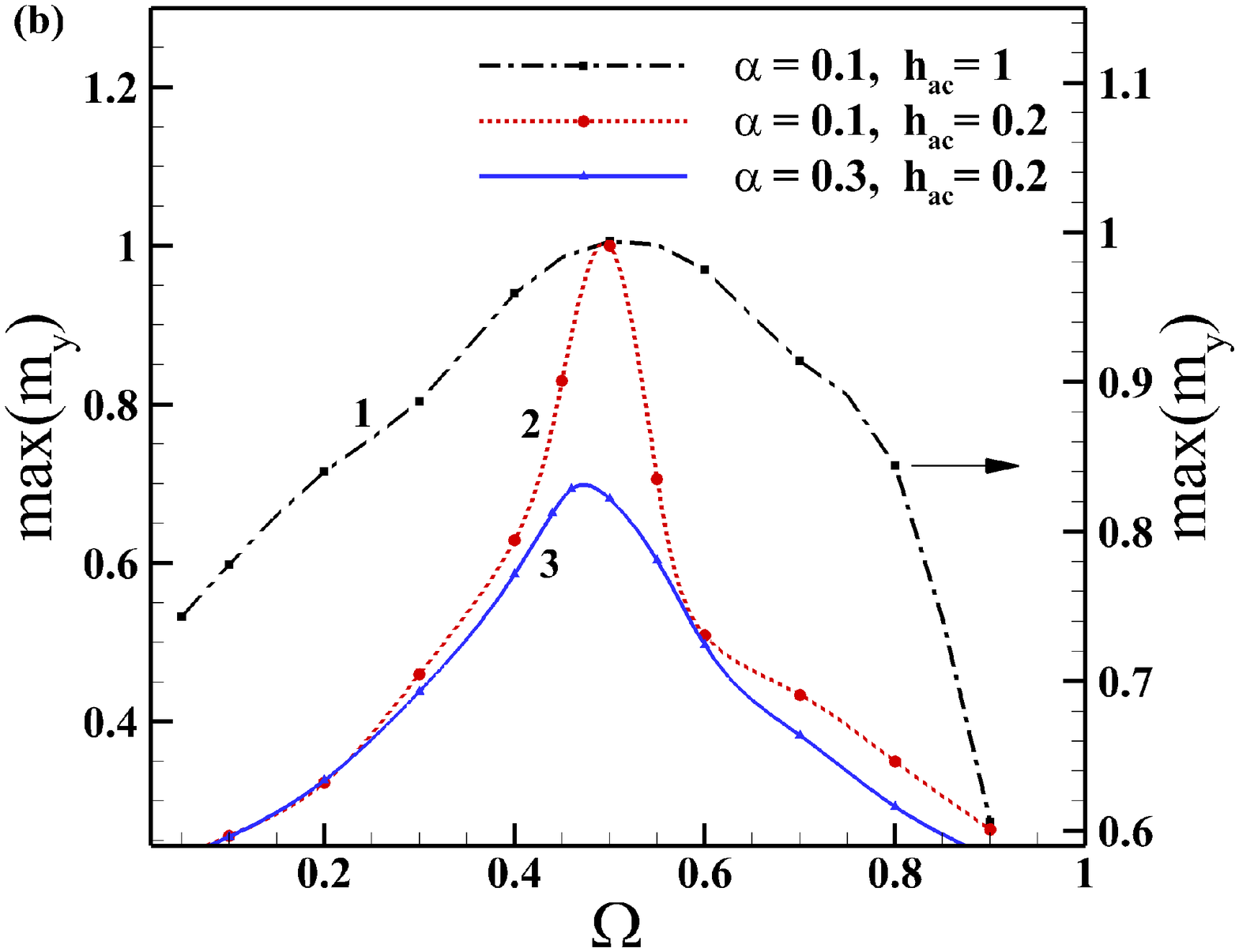}
\includegraphics[width=.83\linewidth]{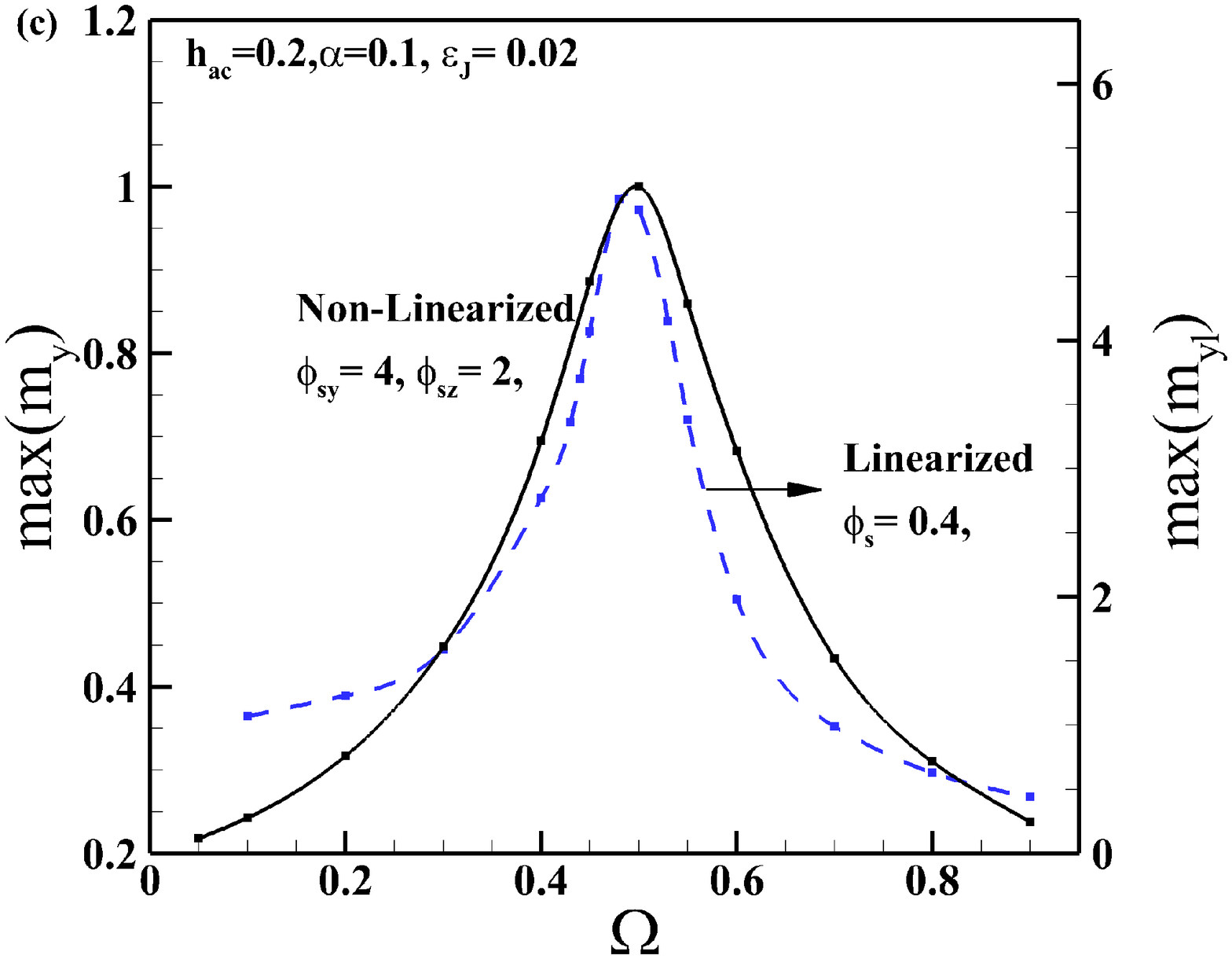}
	\caption {(Color online) (a) Manifestation of the FMR in the frequency dependence of the maximum of magnetization component $m_y$  and the average critical current density  at bias current $I=1.16$. Lines added to guide the eye; (b) Frequency dependence of the maximum of magnetization component $m_y$ at different damping $\alpha$ and amplitude of circularly polarized magnetic field $h_{ac}$. Other parameters are the same as in (a); (c) Comparison with a linearized case at $\epsilon_{J}=0.02$. \label{fig2}}
\end{figure}

First we show that the system displays ferromagnetic resonance. Its manifestation, in the frequency dependence
of the amplitude of the magnetization component $m_y$ and the average critical current density, is presented in Fig.~\ref{fig2}, where we see that the maximum in both cases occurs at the resonance frequency $\Omega=\Omega_{0}=0.5$. Furthermore, the oscillation amplitude is not symmetric relative to $\Omega_{0}$, which reflects the influence of $H_s$ in the effective field. The behavior of the amplitude of the magnetization component $m_x$ is qualitatively the same.

In Fig.~\ref{fig2} (b) we show the frequency dependence of the maximum of magnetization component $m_y$ at different damping $\alpha$ and amplitude of circularly polarized magnetic field $h_{ac}$. We see that the resonance line width changes with changing $h_{ac}$ (curves with label 1 and 2) and $\alpha$ (curves with label 2 and 3). For comparison, we also demonstrate the manifestation of the ferromagnetic resonance in the linearized case.\cite{supplemental,gurevich1996magnetization} In the linearized case, the RSJ equation reduced to
\begin{equation}
I/I_{c} = \frac{ \sin \left(\phi_{s} m_{y}\right)}{ (\phi_{s} m_{y})} \sin \theta (t)  +  \frac{d\theta(t)}{dt},
\label{eq:linearize:RSJ}
\end{equation}
where $I_{c}=I^{0}_{c}\sin(\phi_{sy})/(\phi_{sy})$, $\phi_{sy}=4\pi^2 L_{y} d  M_{z}/\Phi_{0}$
and the expression for $y$-component of magnetization has a form
\begin{equation}
m_{y}=   \frac{- 2  \alpha \frac{\Omega^{2}}{\Omega_{0}^{2}} \cos (\Omega t) + \left(1-\eta_{1} \frac{\Omega^{2}}{\Omega_{0}^{2}}\right) \sin (\Omega t)}{\left( 1-\eta_{2} \frac{\Omega^{2}}{\Omega_{0}^{2}}\right) ^{2} +\Delta_{J}\left(1-\eta_{1}  \frac{\Omega^{2}}{\Omega_{0}^{2}}\right)+ 4 \alpha^{2}\frac{\Omega^{2}}{\Omega_{0}^{2}} },
\label{eq:my_lina1}
\end{equation}
where $\Delta_{J}=\epsilon_{J} \phi_{sz}^{2}\cos\theta(t)/3$, $\eta_{1}=1-\alpha^{2}$ and $\eta_{2}=1+\alpha^{2}$.

Results of calculations based of these formulas are presented in Fig.~\ref{fig2}(c). We see a qualitative agreement of the ferromagnetic resonance features in both cases.

\section{DS structure in the IV characteristics}

\begin{figure}[!ht]
	\includegraphics[width=.85\linewidth]{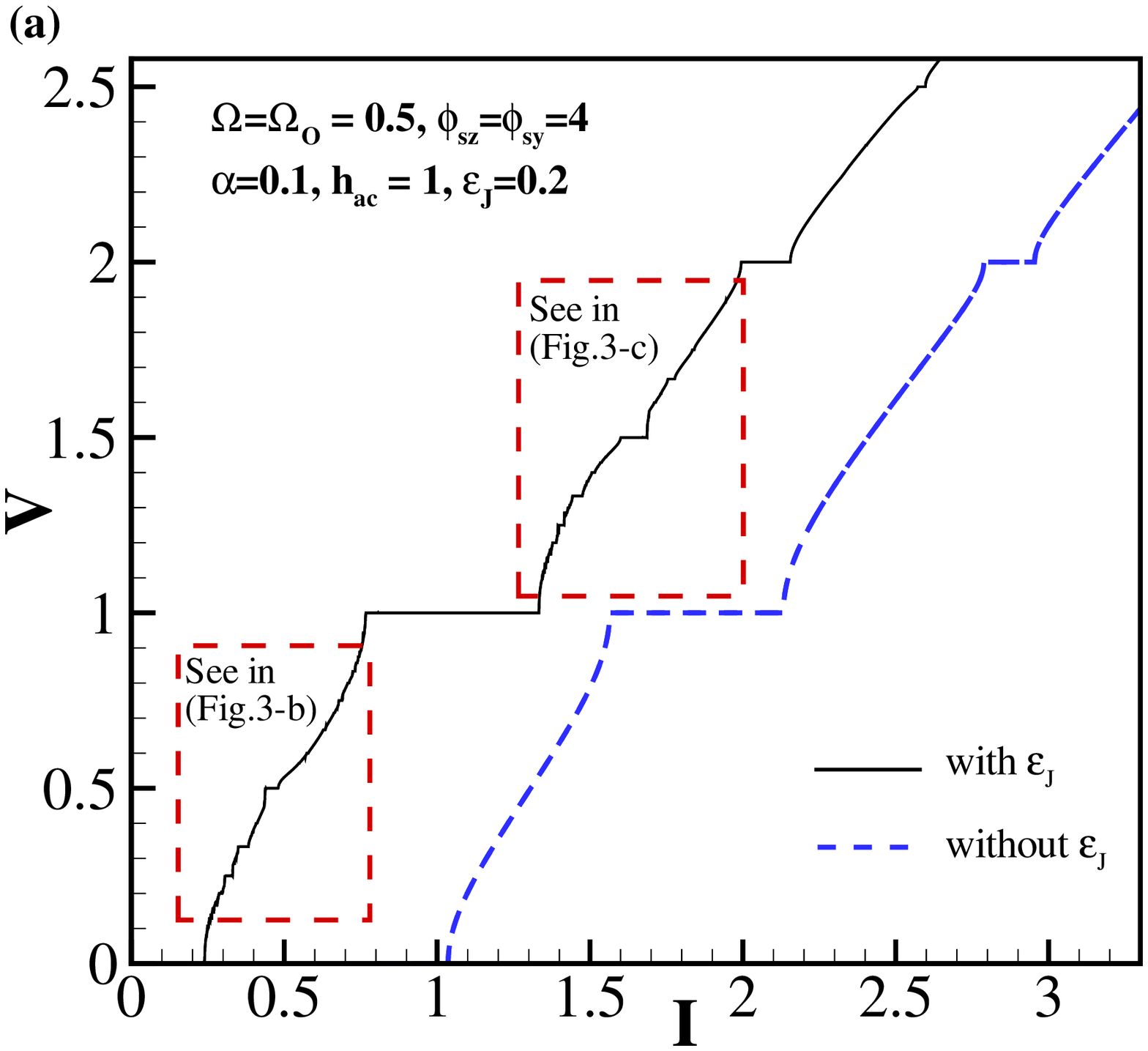}
	\includegraphics[width=.85\linewidth]{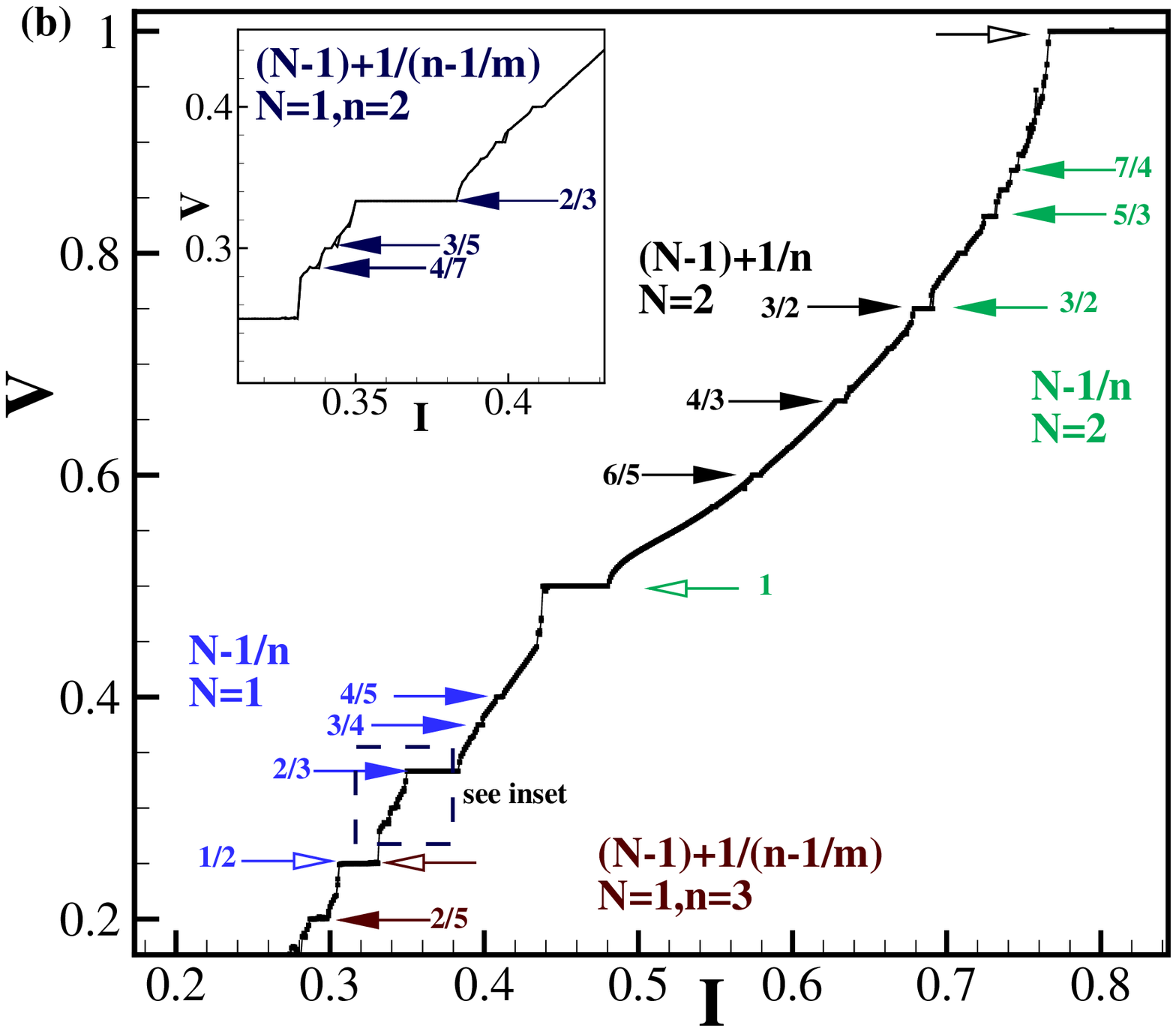}
	\includegraphics[width=.85\linewidth]{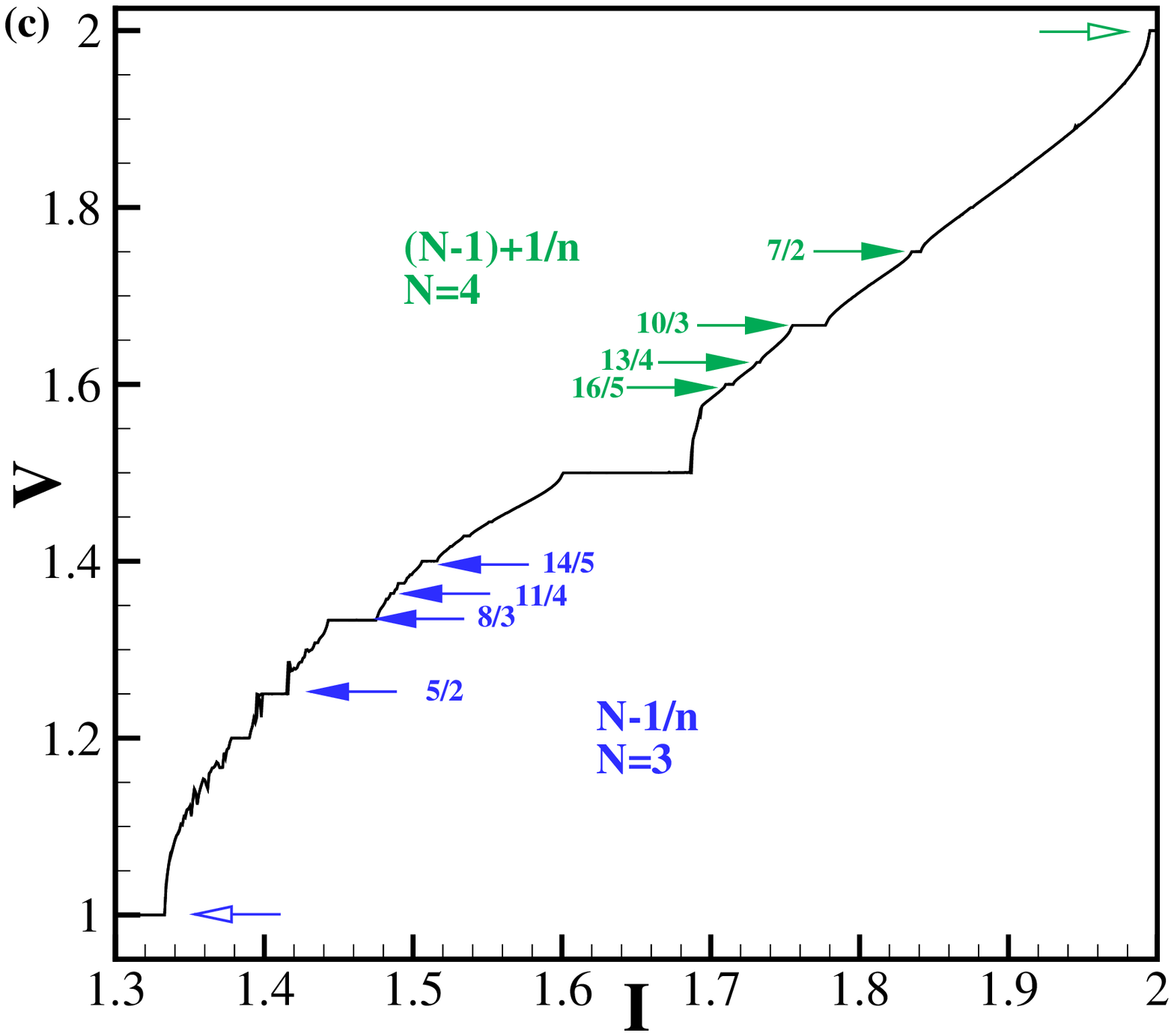}
	\caption{ (Color online) (a) IV characteristic of S/F/S junction at ferromagnetic resonance. The case in Ref.~\onlinecite{hikino2011ferromagnetic} is shown by the dashed line for comparison, shifted by 0.8 to the right for clarity; (b) and (c) enlarge the parts of IV characteristic marked by rectangles in (a). \label{fig3}}	
\end{figure}

Let us now discuss the S/F/S junction at FMR, when the coupling between Josephson and magnetic system is strongest. In Fig.~\ref{fig3}(a)  the IV characteristic demonstrates current steps at $V=m\Omega_0$, with $m$  integer, and also some fractional steps.

In the case of conventional JJs the widths of the first Shapiro step is larger than the second.  In the present case, we see that the width of the first step is much narrower than that of the second. So, the width of the harmonics are different for even and odd $m$: large steps are at even $m$ and small steps at odd $m$. In Ref.~\onlinecite{hikino2011ferromagnetic}, which did not consider the Josephson energy in the expression for the effective field, only the steps with even $m$ were observed. In our case, taking into account the Josephson energy in the effective field, we have obtained additional steps with odd and fractional values of $m$, as we see in Fig.~\ref{fig3}(a).

The structure of those fractional steps can be clarified by analysis of their positions on the voltage scale, using an algorithm based on the generalized continued fraction formula:
\cite{shukrinov2013devil,shukrinov2014structured,sokolovic2017devil}
\begin{equation}
V=\left( N \pm \frac{1}{n\pm
	\frac{1}{m\pm\frac{1}{p\pm..}}}\right)\Omega,
\label{eq8}
\end{equation}
where $N$, $n$, $m$, $p$, $\ldots$ are positive integers. The locking
of the Josephson frequency to the frequency of magnetic precession occurs due to the additional terms ($\Gamma_{yz} \epsilon_{J} \cos \theta,\Gamma_{zy} \epsilon_{J} \cos \theta  $) in the effective field, as given by
Eq.~(\ref{eq6}). Fig.~\ref{fig3}(b) and (c) demonstrate the
enlarged parts of the IV characteristic shown in Fig.~\ref{fig3}(a). There are the fractional current steps between $V=0$ and $V=0.5$ which can be described by the continued fractions of second level\cite{shukrinov2013devil} $(N-1)+1/n$  and $N-1/n$ with $N=1$ in both cases (see Fig.~\ref{fig3}(b)).  In addition, there is  a manifestation of two third-level continued fractions $(N-1)+1/(n-1/m)$ with $N=1$, $n=2$ (shown in the inset) and $n=3$. The steps  between $V=0.5$ and $V=1$ follow the continued fractions of second level $(N-1)+1/n$ and $N-1/n$ with $N=2$ in both cases. In Fig.~\ref{fig3}(c) we see clearly the manifestation of
second level continued fractions $N-1/n$ with $N=3$ and $(N-1)+1/n$ with $N=4$  between voltage steps $V=1$ and $V=2$.

\section{Effect of oscillating electric field}

\begin{figure}[!ht]
	\includegraphics[width=.9\linewidth]{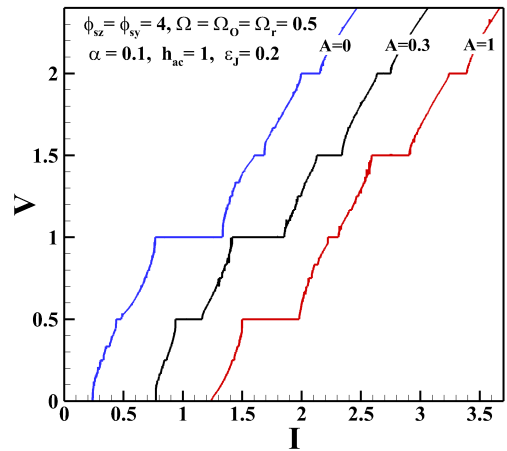}
	\caption{(Color online) IV characteristics of the S/F/S junction at ferromagnetic resonance without oscillating electric field ($A=0$) and two characteristics at amplitudes $A=0.3$ and $A=1$. Here $\Omega_r=\omega_r/\omega_c$, other parameters are the same as in Fig.~\ref{fig3}.  For clarity, the curves at $A=0.3$ and $A=1$ have been shifted to the right, by $\Delta I= 0.6$ and $\Delta I= 1.2$ respectively, relative to the IV characteristic at $A=0$.  \label{fig4}}
\end{figure}

The ac field can affect the Josephson junction directly, and not only
via the oscillating magnetization. The effect of an oscillating electric field from microwave radiation is usually taken into account by adding the term $A \sin \Omega_{r} t$ in Eq.~(\ref{eq1}), where $A$ is the amplitude and $\Omega_r=\omega_{r}/\omega_{c}$ - the frequency of the external electromagnetic radiation. Figure~\ref{fig4} shows the IV characteristics without the effect of the oscillating electric field (i.e. for $A=0$) and two curves at amplitudes $A=0.3$ and $A=1$. In comparison to $A=0$, where the width of the first step at $V=0.5$ is smaller relatively to the step at $V=1$ (a signature of the S/F/S IV characteristics), we see that at $A=0.3$ the first step has widened in comparison to the second step at $A=1$. But, even in this case, the IV characteristics show the unusual behavior of Shapiro step widths for a conventional Josephson junction, specifying width of odd and even steps.

\section{Discussion}

\begin{figure}[!ht]
	\includegraphics[width=.95\linewidth]{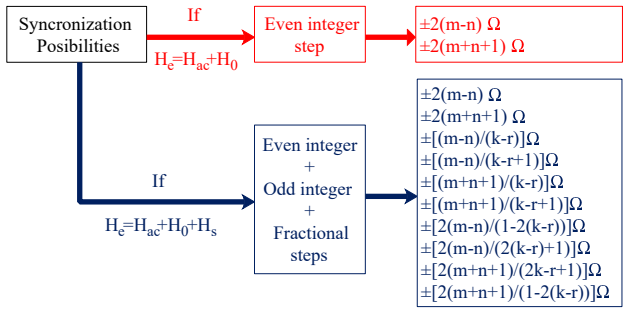}
	\caption{(Color online) Different possibilities of the frequency locking excluding (red) and including (blue) the Josephson energy in the effective magnetic field. \label{fig5}}
\end{figure}

We have also found that one can control the structure of the devil's staircase by tuning the frequency of the ac-magnetic field out of resonance. Of course, the width of the subharmonic steps is largest at the FMR. The step structure depends on the junction parameters (Gilbert
damping, cross-section, etc). The main parameter determining the appearance of the DS structure is the ratio of the Josephson to magnetic energy. If this ratio is close to zero, we observe only even steps. In the present work the appearance of the fractional steps and the formation of the devil's staircases in the IV-curve are consequences of including the Josephson energy in the effective field, i.e. the term $\epsilon_J$ in (\ref{eq6}). We justify this claim by solving the linearized LLG equation analytically using well known mathematical methods.\cite{Oldham2010,Zwillinger,supl} As demonstrated in Fig.~\ref{fig5}, our proposed model shows different possibilities of the frequency locking leading to even, odd and fractional current steps in IV characteristics of S/F/S junction under an external circularly polarized magnetic field. This fact is in an agreement with the results presented in Fig.~\ref{fig3}.

Let us now discuss the possibility of experimentally observing the effects found in this paper. The main parameter which controls the appearance of the current steps is $\epsilon_{J} =E_{J}/\left( v M_{0} H_{0} \right) $. Using typical junction parameters $d=5\,{\rm nm}$, $L_{y}=L_{z}=75 \,{\rm nm}$, critical current $I^{0}_{c}\approx160 \,\mu {\rm A}$, saturation magnetization $M_{0} \approx 4\times 10^{5} \,{\rm A}/{\rm m}$,  $H_{0}\approx26\,{\rm mT}$ and gyromagnetic ration $\gamma=3\pi \,{\rm MHz}/{\rm T}$, we find the value of $\phi_{sy(z)}$=$4\pi^2 L_{y(z)} d M_{0}/\Phi_{0}$ = $3.6$ and $\epsilon_{J}=0.18$, which are very close to the values we used in our simulations and justify the choice  of parameters: $\phi_{sy,sz} =4$, $\epsilon_{J}=0.2$. With the same junction parameters one can control the appearance of the subharmonic steps by tuning the strength of the constant magnetic field $H_{0}$. Estimations show that, for $H_{0}=90 \,{\rm mT}$, the fractional subharmonic steps are disappear at $\epsilon_{J}=0.05$. For junctions with $L_y = L_{z}=50 \, {\rm nm}$, $H_{0}=10 \,{\rm mT}$, we find $\phi_{sy(z)} = 2.4$ and $\epsilon_{J}=1.05$, which are rather good for the step manifestation. Of course, in general, the subharmonic steps are sensitive to junction parameters, Gilbert damping and the frequency of the magnetic field.

\section{Conclusion}
The S/F/S Josephson junction is of considerable importance for the development of certain
spintronic applications/devices. Motivated by physical considerations,
our paper has presented a major advance in modeling the S/F/S Josephson
junction, by including a previously neglected physical effect, i.e.
of the Josephson energy on the effective magnetic field. Our
calculations predict that the addition of the Josephson energy should
manifest itself (measurably) through the appearance of devil's
staircase structures in the IV characteristics, thus providing 
insight into the precise nature of the current-phase relation and 
opportunities for potential applications.

In our paper we have developed a model which fully describes the dynamics of the S/F/S Josephson junction under an applied circularly polarized magnetic field. Manifestation of ferromagnetic resonance in the frequency dependence of the amplitude of the magnetization and the average critical current density was demonstrated. The IV characteristics showed subharmonic steps which formed devil's staircase structures, following the continued fraction algorithm.\cite{shukrinov2013devil} The origin of the found steps was related to the effect of the magnetization dynamics on the phase difference in the Josephson junction. Analytical considerations of the steps were in agreement with the numerical results.

An interesting question appears about whether the manifestation of the staircase structure in the IV characteristics can provide information on the current-phase relation of the S/F/S Josephson junction and, in some cases, serve as a novel method for its determination. The results on the developed model might serve for better understanding of the  coupling between the superconducting current and magnetization in the S/F/S Josephson junction.  The appearance of the staircase structure in experimental situations and its  connection with the current-phase relation may open horizons in this field. The observed features might also find application in some fields of superconducting spintronics.

\section*{Acknowledgments}

The authors thank A. Buzdin, S. Maekawa, S. Takahashi, S. 
Hikino, I. Rahmonov, K. Kulikov, I. Bobkova, and A. Bobkov 
for useful discussions, and D. Kamanin, V. V. Voronov, and 
H. El Samman for supporting this work. The reported study 
was partially funded by RFBR according to the research project 
18-02-00318, and the SA-JINR and Egypt-JINR collaborations. Y.M.S. and A.E.B. thank the visiting researcher program 
at the University of South Africa for financial support.

\newpage
\clearpage
\newpage
\newpage
\begin{widetext}
	\section*{	Supplemental Material to ``Devil's staircases in the IV characteristics of superconductor/ferromagnet/superconductor Josephson junctions''}
	\section{Linearized Landau-Lifshitz-Gilbert equation}
	\subsection{Magnetization Dynamics}
	The Landau-Lifshitz-Gilbert (LLG) equation  for the superconductor/ferromagnet/ superconductor (S/F/S) structure describes the behavior of magnetization in the effective magnetic field $\bm{H}_e$. Being nonlinear, it cannot in general be solved analytically~\cite{LP}.  Here we investigate the system of equations describing the SFS Josephson junction under the application of a circularly polarized magnetic field in $xy$-plane. We derive a linearized form of the equations by using the method of complex amplitudes~\cite{gurevich1996magnetization} and find expressions for the magnetization components. After that, we calculate the effective field components and obtain an expression for $m_{y}$, which we will subsequently use in the RSJ equation.
	
	In its general form, the Landau-Lifshitz-Gilbert equation reads
	\begin{eqnarray}
	\dfrac{d\bm{M}}{dt}=-\gamma\bm{M}\times\bm{H}_e+\dfrac{\alpha}{\lvert\bm{M}\lvert}\left(\bm{M}\times \frac{d\bm{M}}{dt} \right),  \label{eq:LLG}
	\end{eqnarray}
	where $\alpha$ is the Gilbert damping and $\gamma$ is the gyromagnetic ratio.  We assume that the effective magnetic field and magnetization can be written as sums of constant and alternating parts
	\begin{equation}
	\bm{H}_{e}=\bm{H}_{0}+\bm{\tilde{H}}, \ \ \  \ \ \bm{M}=\bm{M}_{0} + \bm{\tilde{M}},
	\label{eq:stead_alter}
	\end{equation}
	where the components of $\bm{H}_{0}$ are $(0,0,H_{0})$,  with $H_{0}=\omega_{0}/\gamma$ and $\omega_{0}$ is the ferromagnetic resonance frequency. The components of $\bm{\tilde{H}}$ are  $(\tilde{H}_{x}, \tilde{H}_{y},0)$. On other hand, the components of $\bm{M}_{0}$ are $ (0,0,M_{z})$ and those of $\bm{\tilde{M}}$ are $(\tilde{M}_{x}, \tilde{M}_{y}, 0)$. The magnitude of alternating parts are considered smaller than the steady parts, i.e.  $\tilde{H} << H_{0} $, $\tilde{M}<< M_{z}$. The linearization of Eq.~(\ref{eq:LLG}) can be found by inserting Eq.~(\ref{eq:stead_alter}) into Eq.~(\ref{eq:LLG}) and neglecting the products of the alternating parts. This gives
	\begin{equation}
	\frac{d\bm{\tilde{M}}}{dt} +\gamma \bm{\tilde{M}} \times \bm{H}_{0}+\frac{\alpha}{\rvert\bm{M}\rvert} \left(\frac{d\bm{\tilde{M}}}{dt}\times \bm{M}_{0}\right)= -\gamma \bm{M}_{0}\times\bm{\tilde{H}}.
	\label{eq:LinLLG1}
	\end{equation}
	We consider a harmonic time dependence for $\bm{\tilde{H}}$. In this case, the time dependence of $\bm{\tilde{M}}$ will be also harmonic. Our aim is to find harmonic solutions to the linearized LLG equation, i.e. in the form
	\begin{equation}
	\bm{\tilde{M}}= \textit{\textbf{\text{\~{m}}}} \ e^{i\omega t}, \ \ \ \ \ \ \bm{\tilde{H}} = \textit{\textbf{\text{\~{h}}}} \ e^{i\omega t}.
	\label{eq:comp_l}
	\end{equation}
	This can be done by inserting Eq.~(\ref{eq:comp_l}) into  Eq.~(\ref{eq:LinLLG1}) to obtain
	\begin{equation}
	i \omega \textit{\textbf{\text{\~{m}}}} +\gamma \textit{\textbf{\text{\~{m}}}} \times \bm{H}_{0}+\frac{i \omega \alpha}{\rvert\bm{M}\rvert}  \textit{\textbf{\text{\~{m}}}}\times \bm{M}_{0}= -\gamma \bm{M}_{0}\times\textit{\textbf{\text{\~{h}}}}.
	\label{eq:LinLLG2}
	\end{equation}
	Projecting Eq.~(\ref{eq:LinLLG2}) onto the axes of Cartesian coordinate system, we get
	\begin{eqnarray}
	i \omega \textit{\text{\~{m}}} _{x}+(\omega_{0}+i \alpha \omega) \textit{\text{\~{m}}} _{y}&=&\gamma M_{z} \textit{\text{\~{h}}}_{y},\nonumber \\
	-(\omega_{0}+i \alpha \omega) \textit{\text{\~{m}}} _{x}+i \omega\textit{\text{\~{m}}} _{y}&=&-\gamma M_{z}\textit{\text{\~{h}}}_{x},
	\label{eq:eqofmotion}
	\end{eqnarray}
	where we use $|\bm{M}|\approx M_{z}$, $H_{0}=\omega_{0}/\gamma$. The solution to Eq.~(\ref{eq:eqofmotion}) is
	\begin{eqnarray}
	\textit{\text{\~{m}}} _{x}  &=& \frac{(\omega_{0}+i\alpha \omega) \gamma M_{z} \textit{\text{\~{h}}}_{x}+i \omega \gamma M_{z} \textit{\text{\~{h}}}_{y}}{\omega_{0}^{2}-(1+\alpha^{2})\omega^{2}+2i\alpha \omega\omega_{0}}, \nonumber \\
	\textit{\text{\~{m}}} _{y}  &=&  \frac{-i \omega \gamma M_{z} \textit{\text{\~{h}}}_{x} +(\omega_{0}+i\alpha \omega) \gamma M_{z} \textit{\text{\~{h}}}_{y}}{\omega_{0}^{2}-(1+\alpha^{2})\omega^{2}+2i\alpha \omega\omega_{0}},  
	\label{eq:sysequ}
	\end{eqnarray}
	$\textit{\text{\~{m}}}_{x}(t)$ and $\textit{\text{\~{m}}}_{y}(t)$ can be written in the following forms
	\begin{eqnarray}
	\textit{\text{\~{m}}}_{x}  &=& \chi_{1}' \textit{\text{\~{h}}}_{x} +\chi '_{2} \textit{\text{\~{h}}}_{y} -  i( \chi_{1} ''\textit{\text{\~{h}}}_{x}- \chi''_{2} \textit{\text{\~{h}}}_{y}), \nonumber \\
	\textit{\text{\~{m}}}_{y}  &=& -  \chi '_{2}\textit{\text{\~{h}}}_{x}  + \chi_{1}'  \textit{\text{\~{h}}}_{y} - i (\chi''_{2} \textit{\text{\~{h}}}_{x} +  \chi_{1} '' \textit{\text{\~{h}}}_{y}),
	\label{eq:sysequ1}
	\end{eqnarray}
	where
	\begin{eqnarray}
	\chi_{1}' &=& \frac{1}{\varGamma} \gamma M_{z} \omega_{z} [\omega_{0}^{2} - (1-\alpha^{2})\omega^{2}], \nonumber \\
	\chi_{1}'' &=& \frac{1}{\varGamma} \alpha \gamma M_{z} \omega [\omega_{0}^{2} + (1+\alpha^{2})\omega^{2}], \nonumber \\
	\chi'_{2} &=& \frac{1}{\varGamma} 2 \alpha \gamma M_{z} \omega^{2} \omega_{0}, \nonumber \\
	\chi''_{2} &=& \frac{1}{\varGamma} \gamma M_{z} \omega [\omega_{0}^{2} - (1+\alpha^{2})\omega^{2}], \nonumber \\
	\varGamma &=& [\omega_{0}^{2} - (1+\alpha^{2})\omega^{2}]^{2} + 4 \alpha^{2} \omega^{2}\omega_{0}^{2}.
	\end{eqnarray}
	Using Eq.(\ref{eq:comp_l}), the real part for $M_{x}(t)$ and $M_{y}(t)$ can be written as
	\begin{eqnarray}
	Re\{M_{x}(t)\} &=&\frac{\gamma M_{z}}{\omega_{0}} \left[ \frac{\left( 1-(1-\alpha^{2}) \frac{\omega^{2}}{\omega_{0}^{2}}\right)  H_{x}(t)+2 \alpha \frac{\omega^{2}}{\omega_{0}^{2}} H_{y}(t)}{\left( 1-(1+\alpha^{2})\frac{\omega^{2}}{\omega_{0}^{2}}\right) ^{2} + 4 \alpha^{2}\frac{\omega^{2}}{\omega_{0}^{2}}}\right],\\
	Re\{M_{y}(t)\} &=& \frac{\gamma M_{z}}{\omega_{0}} \left[ \frac{-2  \alpha \frac{\omega^{2}}{\omega_{0}^{2}} H_{x}(t)+ \left( 1-(1-\alpha^{2}) \frac{\omega^{2}}{\omega_{0}^{2}}\right)  H_{y}(t)}{\left( 1-(1+\alpha^{2})\frac{\omega^{2}}{\omega_{0}^{2}}\right) ^{2} + 4 \alpha^{2}\frac{\omega^{2}}{\omega_{0}^{2}}}\right].
	\label{eq:my_expre}
	\end{eqnarray}

	\subsection{Expression for Effective Field and Magnetization Component}
	Next, we find the effective field components $H_{x}(t)$ and $H_{y}(t)$ which should have harmonic dependence. The total energy of the S/F/S Josephson junction in the circularly polarized ac field is given by $E=E_{s} + E_{M} + E_{ac}$, where
	\begin{eqnarray}
	E_{s} &=& -\frac{\Phi_{0}}{2\pi} \theta(y,z,t) I+ E_{J} [1-\cos(\theta(y,z,t))], \nonumber \\
	E_{M} &=& - v H_{0}  M_{z}, \nonumber \\
	E_{ac}&=& - v M_{x} H_{ac} \cos \omega t - v M_{y} H_{ac} \sin \omega t,
	\end{eqnarray}
	Here $E_{J}=I_{c} \Phi_{0}/2\pi$, $I_{c}$ is the critical current, $\Phi_{0}=h/(2e)$ is the magnetic flux quantum, $\theta(y,z,t)$ is the gauge invariant phase difference between superconducting electrodes, $\theta(y,z,t)=\theta (t)-8 \pi^{2}  d M_z y/\Phi _0+8 \pi^{2}  d M_y(t) z/\Phi _0$, d is the magnetic thickness, and  $v$ is the volume. $H_{ac}$ and  $\omega$ are the amplitude and frequency of the ac magnetic field. The effective field can be found using $\bm{H}_{e}= - \nabla_{\textbf{M}} E/v$.
	In the dimensionless form we use $m=\bm{M}/M_{z}$ (we omit the time argument for simplicity of notation), $\bm{h}_{e} = \bm{H}_{e}/H_{0}$, $h_{ac} =H_{ac}/H_{0}$,  $\Omega = \omega /\omega_{c}$,  $\Omega_{0} = \omega_{0} /\omega_{c}$, and $t\longrightarrow t\omega_{c}$. Here  $\omega_{c}=2 \pi I_{c} R/\Phi_{0}$ is the characteristic frequency. After integrate over the junction area, the effective field components in x and y-direction are given in dimensionless form by
	\begin{eqnarray}
	h_{x}&=&h_{ac} \cos (\Omega t),  \\
	h_{y} &=& h_{ac} \sin (\Omega t)  + \frac{\epsilon_{J} \cos \theta(t)}{m_y}\left(\cos \left(\phi_{sz} m _y\right)-\frac{\sin \left(\phi_{sz} m _y\right)}{\phi_{sz} m _y}\right),
	\label{eq:effectivefield}
	\end{eqnarray}
	where  $\phi_{sz}=4\pi^2 L_{z} d M_{z}/\Phi_{0}$, d is the thickness of ferromagnet, $L_{y},L_{z}$ are the junction lengths in y and z-direction. Since $M_{z}$ is constant, we define $I_{c}= I^{0}_{c} \sin(\phi_{sy} )/(\phi_{sy} )$,  $\epsilon_{J}=\gamma E_{J}/(v M_{z} \Omega_{0})$, and $\phi_{sy}=4\pi^2 L_{y} d  M_{z}/\Phi_{0}$. Since we assume that $\tilde{M}<<M_{z}$, we use series expansion for $\cos (\phi_{sz} m _y)$ and $\sin(\phi_{sz} m _y)$ in the second term of $h_{y}$ (see Eq.~(\ref{eq:effectivefield})). So, we obtain for first order approximation
	\begin{equation}
	\frac{1}{m_{y}} \left(\cos \left(\phi_{sz} m_y\right)-\frac{\sin \left(\phi_{sz} m _y\right)}{\phi_{sz} m_y}\right) \approx  - \frac{\phi_{sz}^{2} m_{y}}{3}+....
	\label{eq:series}
	\end{equation}
	The effective field reads
	\begin{equation}
	h_{y}=-\frac{\epsilon_{J}  \phi_{sz}^{2} m_{y}}{3}  \cos \theta(t)  + h_{ac} \sin \Omega t.
	\label{eq:hy_new}
	\end{equation}
	This term is considered as a modulated harmonic behavior that  depends on the value of  $(\epsilon_{J} \phi_{sz}^{2} m_{y}/3) \cos\theta(t)$. So, we can rewrite $m_{y}$, using Eq.~(\ref{eq:my_expre}) and Eq.~(\ref{eq:hy_new}) in the following form
	\begin{equation}
	m_{y}=   \frac{- 2  \alpha \frac{\Omega^{2}}{\Omega_{0}^{2}} \cos (\Omega t) + \left(1-(1-\alpha^{2}) \frac{\Omega^{2}}{\Omega_{0}^{2}}\right) \sin (\Omega t)}{\left( 1-(1+\alpha^{2})\frac{\Omega^{2}}{\Omega_{0}^{2}}\right) ^{2} +\Delta_{J}\left(1-(1-\alpha^{2}) \frac{\Omega^{2}}{\Omega_{0}^{2}}\right)+ 4 \alpha^{2}\frac{\Omega^{2}}{\Omega_{0}^{2}} },
	\label{eq:my_lina11}
	\end{equation}
	
	where $\Delta_{J}=\epsilon_{J} \phi_{sz}^{2} \cos\theta(t)/3$. At $\alpha=0$ we have:
	\begin{equation}
	m_{y}=  \frac{3 \Omega_{0}^2 \sin (\Omega t )}{3( \Omega_{0}^2- \Omega ^2)+\Omega_{0}^2 \epsilon_{J}\phi_{sz}^2 \cos (\theta(t))}.
	\label{eq:my_alpha_0}
	\end{equation}
	Now, let us describe the result shown by Eq.(\ref{eq:my_lina}). If $\bm{H}_{e}=\bm{H}_{ac}+\bm{H}_{0}$, the oscillation of $m_{y}$ is purely harmonic in time. While if we take Josephson energy in the effective field, the oscillation of $m_{y}$ can be considered as harmonic with small changes due to Josephson energy. The validity of the above method require a harmonic dependence of the effective field which can be satisfied in Eq.(\ref{eq:hy_new}).

	\subsection{RSJ Equation}
	
	According to RCSJ model, the current through the junction in the dimensionless form is given by \cite{stewart1968current,mccumber1968effect}
	\begin{equation}
	I =I^{0}_{c} \sin \theta(y,z,t) + \frac{d\theta(y,z,t)}{dt} + \beta_{c}\frac{d^{2}\theta(y,z,t)}{dt^{2}},
	\label{eq:RSJj}
	\end{equation}
	where $I^{0}_{c}$ is the critical current, $\beta_{c}=RC \omega_{c}$ is the McCumber parameter (Here we consider $\beta_{c}=0$). The Modified RCSJ equation is found by inserting $\theta(y,z,t)=\theta (t)-8 \pi^{2}  d M_z y/\Phi _0+8 \pi^{2}  d M_y(t) z/\Phi _0$ into Eq.(\ref{eq:RSJj}) then take the integration over junction area ($\int _{-\text{L}_z/2}^{\text{L}_z/2}\int _{-\text{L}_y/2}^{\text{L}_y/2} ... dy dz/L_{z}L_{y}$). The final RCSJ equation in the dimensionless form is given by
	
	\begin{equation}
	I/I_{c} = \frac{ \sin \left(\phi_{s} m_{y}\right)}{ (\phi_{s} m_{y})} \sin \theta (t)  +  \frac{d\theta(t)}{dt},
	\label{eq:linearize:RSJq}
	\end{equation}
	where $I_{c}=I^{0}_{c}\sin(\phi_{sy})/(\phi_{sy})$, $\phi_{sy}=4\pi^2 L_{y} d  M_{z}/\Phi_{0}$,  and $\phi_{s}=4\pi^2 L_{z} d M_{z} h_{ac}/\Phi_{0} $. In the next section, we will use the expression for $m_{y}$ given by Eq.(\ref{eq:my_lina11}) in the supercurrent term in the RSJ equation to find the conditions for the appearance of current steps.

	\subsection{Origin of Current Steps in the IV characteristics}
	
	As we stress in the main text, adding the part corresponding to the Josephson energy in the effective field, leads to the appearance of subharmonic Shapiro-like steps in IV characteristics of our system. Here we demonstrate the origin of such current steps by analytical considerations. To this end we first analyze the supercurrent term $I_{s}=I^{0}_{c} \sin(\theta(y,z,t))$. The gauge invariant phase difference between superconducting electrodes is given by
	\begin{equation}
	\theta(y,z,t)=\theta (t)-\frac{8 \pi^{2}  d M_z } {\Phi _0} y+\frac{8 \pi^{2}  d M_y(t)}{\Phi _0}z
	\end{equation}
	where $ \theta(t)=2e  V_{0} t/\hbar  + \theta_{0}=\Omega_{J} t + \theta_{0}$, $\Omega_{J}$ is the Josephson frequency. The supercurrent term in the linearized case  $I_{s}=I^{0}_{c} \sin(\theta(y,z,t))$, is given by (see Eq.~(\ref{eq:linearize:RSJq}))
	\begin{equation}
	I_{s}/I_{c}=\frac{ \sin \left(\phi_{s} m_{y}\right)}{ (\phi_{s} m_{y})} \sin (\Omega_{J} t + \theta_{0}).
	\end{equation}
	where
	\begin{equation}
	m_{y}=  \frac{1}{\Delta}\left[ \frac{-\varUpsilon_{1} \cos (\Omega t) + \varUpsilon_{2} \sin (\Omega t)}{D}\right],
	\label{eq:my_lina}
	\end{equation}
	with
	\begin{eqnarray}
	\varUpsilon_{1}&=& 2  \alpha \frac{\Omega^{2}}{\Omega_{0}^{2}},\nonumber \\
	\varUpsilon_{2}&=& \left(1-(1-\alpha^{2}) \frac{\Omega^{2}}{\Omega_{0}^{2}}\right),\nonumber \\
	D&=&\left( 1-(1+\alpha^{2})\frac{\Omega^{2}}{\Omega_{0}^{2}}\right) ^{2} + 4 \alpha^{2}\frac{\Omega^{2}}{\Omega_{0}^{2}},\nonumber \\
	\Delta&=& 1+ \frac{\phi_{sz}^{2} \epsilon_{J}  \varUpsilon_{2}}{3 D } \cos \theta(t).
	\end{eqnarray}	
	We rewrite $m_{y}$ in Eq.~(\ref{eq:my_lina}) as
	\begin{equation}
	m_{y}= \frac{R}{\Delta} \sin(\Omega t - \zeta),
	\label{eq:my_simplified}
	\end{equation}
	where $R=\frac{1}{D}\sqrt{\varUpsilon_{1}^{2}+\varUpsilon_{2}^{2}}$, $\zeta=\arctan(\varUpsilon_{2}/\varUpsilon_{1})$, $\Delta=1+\xi \cos (\Omega_{J} t + \theta_{0})$ and $\xi=\phi_{sz}^{2} \varUpsilon_{2} \epsilon_{J}/3D$.
	Substituting this expression to the formula (\ref{eq:linearize:RSJq}), we get
	\begin{equation}
	I_{s}=\frac{\sin(\phi_{s} R \sin(\Omega t-\zeta)/\Delta)}{(\phi_{s} R \sin(\Omega t-\zeta)/\Delta)} \sin (\Omega_{J} t + \theta_{0}).
	\label{eq:Is_linar_m2}
	\end{equation}
	Using the following series expansions~\cite{Oldham2010}:
	\begin{eqnarray}
	\sin(\phi_{s} R \sin(\Omega t-\zeta)/\Delta)&=& \frac{2 \Delta}{\phi_{s} R} \sum_{n=0}^{\infty} J_{2n+1} \left(\frac{\phi_{s} R}{\Delta}\right) \sin ((2n+1)(\Omega t-\zeta)), \nonumber \\
	\csc (\Omega t -\zeta) &=& 2 \sum_{m=0}^{\infty} \sin ((2m+1)(\Omega t -\zeta)),
	\end{eqnarray}
	where $J_{2n+1}(\phi_{s} R/\Delta)$ are Bessel functions of the first kind,
	\begin{equation}
	J_{2n+1}\left( \frac{\phi_{s} R}{\Delta}\right)=\sum_{p=0}^{\infty} \frac{(-1) ^{p}}{2^{2p+2n+1} \Gamma(p+2n+2) (2n+1)!} \left( \frac{\phi_{s} R}{\Delta}\right)^{2p+2n+1}.
	\label{eq:bessel}
	\end{equation}
	The expression for the supercurrent then becomes
	\begin{equation}
	I_{s}=\frac{2 \Delta  \sin (\Omega_{J} t + \theta_{0})}{\phi_{s} R} \sum_{m=0}^{\infty} \sum_{n=0}^{\infty}  J_{2n+1}  \left(\frac{\phi_{s} R}{\Delta}\right) [ \cos (2(n-m)(\Omega t-\zeta))- \cos (2(n+m+1)(\Omega t-\zeta))],
	\label{eq:Is_linar2_m2}
	\end{equation}
	So, substituting Eq.~(\ref{eq:bessel}) into Eq.~(\ref{eq:Is_linar2_m2}), we have
	\begin{eqnarray}
	I_{s}&=&\frac{2 \sin (\Omega_{J} t + \theta_{0})}{\phi_{s} R} \hat{\sum_{m,n,p}}  \left( \frac{1}{1+\xi \cos(\Omega_{J} t + \theta_{0})}\right)^{2p+2n}  \bigg[ \cos (2(n-m)(\Omega t-\zeta))\nonumber
	\\ &-& \cos (2(n+m+1)(\Omega t-\zeta))\bigg],
	\label{eq:Is}
	\end{eqnarray}
	where
	\begin{eqnarray}
	\hat{\sum_{m,n,p}}= \sum_{m=0}^{\infty} \sum_{n=0}^{\infty} \sum_{p=0}^{\infty} \frac{(-1) ^{p} (\phi_{s} R)^{2p+2n+1}}{2^{2p+2n+1} \Gamma(p+2n+2) (2n+1)!}.
	\end{eqnarray}
	Using the binomial expansion~\cite{Oldham2010}
	\begin{equation}
	(1+\xi \cos(\Omega_{J} t + \theta_{0}))^{-2n-2p}=\sum_{k=0}^{\infty} \frac{ (2n+2p)_{k}}{k!} (\xi \cos(\Omega_{J} t + \theta_{0}))^{k}, \ \  |\zeta \cos(\Omega_{J} t + \theta_{0})| <1,
	\label{eq:binomial}
	\end{equation}
	and trigonometric power formulas~\cite{Zwillinger}
	\begin{eqnarray}
	\cos^{2k} (\Omega_{J} t + \theta_{0})&=&\frac{(2k)!}{2^{2k}(k!)^{2}} + \frac{1}{2^{2k-1}} \sum_{r=0}^{k-1} \frac{(2k)!}{(2k-r)! r!}\cos ((2(k-r)(\Omega_{J} t + \theta_{0}))), \nonumber \\
	\cos^{2k+1} (\Omega_{J} t + \theta_{0})&=&\frac{1}{4^{k}}\sum_{r=0}^{k} \frac{(2k+1)!}{(2k-r+1)! r!}\cos ((2k-2r+1)(\Omega_{J} t + \theta_{0})).
	\label{eq:powersum}
	\end{eqnarray}
	we can rewrite Eq.~(\ref{eq:binomial}) in terms of even and odd powers as
	\begin{eqnarray}
	(1+\xi \cos(\Omega_{J} t + \theta_{0}))^{-2n-2p}&=&\sum_{k=0}^{\infty} \frac{ (2n+2p)_{2k}}{(2k)!} (\xi \cos(\Omega_{J} t + \theta_{0}))^{2k} \nonumber \\ &+& \sum_{k=0}^{\infty} \frac{ (2n+2p)_{2k+1}}{(2k+1)!} (\xi \cos(\Omega_{J} t + \theta_{0}))^{2k+1},
	\label{eq:binomial2}
	\end{eqnarray}
	where the Pochhammer symbol $(2n+2p)_{k}=\Gamma (2n+2p+k)/\Gamma (2n+2p)$.  By inserting Eq.(\ref{eq:powersum}) and Eq.(\ref{eq:binomial2}) in Eq.(\ref{eq:Is}), we obtain
	\begin{eqnarray}
	I_{s}&=&\frac{2 }{\phi_{s} R} \hat{\sum_{m,n,p}}
	[ \cos (2(n-m)(\Omega t-\zeta))- \cos (2(n+m+1)(\Omega t-\zeta))] \nonumber \\ & &\bigg( \sum_{k=0}^{\infty} \frac{ \xi^{2k}  (2n+2p)_{2k}}{(2k)!} \bigg[ \frac{(2k)!}{2^{2k}(k!)^{2}} + \frac{1}{2^{2k-1}} \sum_{r=0}^{k-1} \frac{(2k)!}{(2k-r)! r!}\cos ((2(k-r)(\Omega_{J} t + \theta_{0}))) \bigg]\nonumber \\  &+&  \sum_{k=0}^{\infty} \frac{\xi^{2k+1} (2n+2p)_{2k+1}}{(2k+1)!}  \bigg[ \frac{1}{4^{k}}\sum_{r=0}^{k} \frac{(2k+1)!}{(2k-r+1)! r!}\cos ((2k-2r+1)(\Omega_{J} t + \theta_{0}))\bigg] \sin (\Omega_{J} t + \theta_{0}) \bigg). \nonumber \\
	\label{eq:Is_1}
	\end{eqnarray}
	Finally, after using the trigonometry relation $\sin A\cos B = (1/2) [\sin (A+B) + \sin(A-B)]$, the supercurrent can be expressed as
	{\footnotesize
		\begin{eqnarray}
		I_{s}&=&\frac{1 }{\phi_{s} R} \hat{\sum_{m,n,p}}
		\bigg[ \sum_{k=0}^{\infty} \frac{\xi^{2k}  (2n+2p)_{2k}}{2^{2k+1}(k!)^{2}} \big\lgroup \sin[(\Omega_{J} t + \theta_{0})+2(n-m)(\Omega t-\zeta)] +\sin[(\Omega_{J} t + \theta_{0})-2(n-m)(\Omega t-\zeta)] \nonumber \\ & -&\sin[(\Omega_{J} t + \theta_{0})+2(n+m+1)(\Omega t-\zeta)]- \sin[(\Omega_{J} t + \theta_{0})-2(n+m+1)(\Omega t-\zeta)]  \big\rgroup\nonumber \\ & +&\sum_{k=0}^{\infty} \sum_{r=0}^{k-1} \frac{\xi^{2k}  (2n+2p)_{2k}}{2^{2k}(2k-r)! r!} \big\lgroup \sin[(2(k-r)+1)(\Omega_{J} t + \theta_{0})+2(n-m)(\Omega t-\zeta)]
		\nonumber \\ & +& \sin[(2(k-r)+1)(\Omega_{J} t + \theta_{0})-2(n-m)(\Omega t-\zeta)]- \sin[(2(k-r)+1)(\Omega_{J} t + \theta_{0})+2(n+m+1)(\Omega t-\zeta))]\nonumber \\ &-&\sin[(2(k-r)+1)(\Omega_{J} t + \theta_{0})-2(n+m+1)(\Omega t-\zeta)]
		+\sin[(1-2(k-r))(\Omega_{J} t + \theta_{0})+2(n-m)(\Omega t-\zeta)]\nonumber \\ &+& \sin[(1-2(k-r))(\Omega_{J} t + \theta_{0})-2(n-m)(\Omega t-\zeta)]- \sin[(1-2(k-r))(\Omega_{J} t + \theta_{0})+2(n+m+1)(\Omega t-\zeta)]\nonumber \\ &-&\sin[(1-2(k-r))(\Omega_{J} t + \theta_{0})-2(n+m+1)(\Omega t-\zeta)]
		\big\rgroup \nonumber \\  &+&  \sum_{k=0}^{\infty}\sum_{r=0}^{k} \frac{\xi^{2k+1} (2n+2p)_{2k+1}}{ 4^{k} 2(2k-r+1)! r! }  \big\lgroup \sin[2(k-r+1)(\Omega_{J} t + \theta_{0})+2(n-m)(\Omega t-\zeta)]\nonumber \\ &+&\sin[2(k-r+1)(\Omega_{J} t + \theta_{0})-2(n-m)(\Omega t-\zeta)]- \sin[2(k-r+1)(\Omega_{J} t + \theta_{0})+2(n+m+1)(\Omega t-\zeta)]\nonumber \\ &-&\sin[2(k-r+1)(\Omega_{J} t + \theta_{0})-2(n+m+1)(\Omega t-\zeta)]-\sin[2(k-r)(\Omega_{J} t + \theta_{0})-2(n-m)(\Omega t-\zeta)]\nonumber \\ &-&\sin[2(k-r)(\Omega_{J} t + \theta_{0})+2(n-m)(\Omega t-\zeta)]+ \sin[2(k-r)(\Omega_{J} t + \theta_{0})-2(n+m+1)(\Omega t-\zeta)]\nonumber \\ &+&\sin[2(k-r)(\Omega_{J} t + \theta_{0})+2(n+m+1)(\Omega t-\zeta)]
		\big\rgroup  \bigg].
		\label{eq:IS_general}
		\end{eqnarray}}
	A special case occurs if one consider $\bm{H}_{e} =\bm{H}_{0}+\bm{H}_{ac}$, where $\bm{H}_{0}=(0,0,\omega_{0}/\gamma)$,  $\mathbf{H}_{ac}= (H_{ac} \cos \omega t, H_{ac} \sin \omega t, 0 )$ represents circularly polarized magnetic field in the xy-plane with amplitude $H_{ac}$. In this case $\Delta=1$ and Eq.~(\ref{eq:Is_linar2_m2}) become
	\begin{eqnarray}
	I_{s}&=&\frac{ 1}{\phi_{s} R} \sum_{m=0}^{\infty} \sum_{n=0}^{\infty}  J_{2n+1}  \left(\phi_{s} R\right) [ \sin[ \Omega_{J} t + \theta_{0}+2(n-m)(\Omega t-\zeta)]+\sin[ \Omega_{J} t + \theta_{0}-2(n-m)(\Omega t-\zeta)]\nonumber \\ &-& \sin [ \Omega_{J} t + \theta_{0}+2(n+m+1)(\Omega t-\zeta)]- \sin [ \Omega_{J} t + \theta_{0}-2(n+m+1)(\Omega t-\zeta)]].
	\label{eq:Is_spcase1}
	\end{eqnarray}
	In Eq.~(\ref{eq:Is_spcase1}) we see four different possibilities, which lead to the appearance of time independent terms in the current when
	\begin{eqnarray}
	\Omega_{J}=\pm2(n-m)\Omega \Longrightarrow I^{1}_{s} &=&\frac{ 1}{\phi_{s} R} \sum_{m=0}^{\infty} \sum_{n=0}^{\infty}  J_{2n+1}  \left(\phi_{s} R\right) \sin ( \theta_{0}\pm2(n-m)\zeta),\nonumber \\
	\Omega_{J}=\pm2 (n+m+1)\Omega \Longrightarrow I^{2}_{s} &=&\frac{ 1}{\phi_{s} R} \sum_{m=0}^{\infty} \sum_{n=0}^{\infty}  J_{2n+1}  \left(\phi_{s} R\right) \sin ( \theta_{0}\pm2(n+m+1)\zeta),
	\end{eqnarray}
	where  $\zeta=\arctan(\varUpsilon_{2}/\varUpsilon_{1})$. Now we apply the above method for the expression of supercurrent in Eq.~(\ref{eq:IS_general}). Time independent terms appear in the current when
	\begin{eqnarray}
	\Omega_{J}&=&\pm2(n-m)\Omega \Longrightarrow I^{1}_{s}=\frac{1 }{\phi_{s} R} \varSigma_{1} \sin[\theta_{0}\pm 2(n-m)\zeta],  \nonumber \\
	\Omega_{J}&=&\pm2(n+m+1)\Omega \Longrightarrow I^{2}_{s}=\frac{1 }{\phi_{s} R} \sin[\theta_{0}\pm 2(n+m+1)\zeta], \nonumber \\
	\Omega_{J}&=&\pm [2(n-m)/(2(k-r)+1)]\Omega \Longrightarrow I^{3}_{s}=\frac{1 }{\phi_{s} R} \varSigma_{2}\sin[(2(k-r)+1)\theta_{0}\pm 2(n-m)\zeta],  \nonumber \\
	\Omega_{J}&=&\pm [2(n+m+1)/(2(k-r)+1)]\Omega \Longrightarrow I^{4}_{s}=\frac{1 }{\phi_{s} R} \varSigma_{2}\sin[(2(k-r)+1)\theta_{0}\pm 2(n+m+1)\zeta], \nonumber \\
	\Omega_{J}&=&\pm [2(n-m)/(1-2(k-r))]\Omega \Longrightarrow I^{5}_{s}=\frac{1 }{\phi_{s} R} \varSigma_{2}\sin[1-2(k-r)\theta_{0}\pm 2(n-m)\zeta],  \nonumber \\
	\Omega_{J}&=&\pm [2(n+m+1)/(1-2(k-r))]\Omega\Longrightarrow I^{6}_{s}=\frac{1 }{\phi_{s} R} \varSigma_{2} \sin[(1-2(k-r)\theta_{0}\pm 2(n+m+1)\zeta],  \nonumber \\
	\Omega_{J}&=&\pm [(n-m)/(k-r+1)]\Omega \Longrightarrow I^{7}_{s}=\frac{1 }{\phi_{s} R}\varSigma_{3} \sin[2(k-r+1)\theta_{0}\pm2(n-m)\zeta],  \nonumber \\
	\Omega_{J}&=&\pm [(n+m+1)/(k-r+1)]\Omega  \Longrightarrow I^{8}_{s}=\frac{1 }{\phi_{s} R} \varSigma_{3}\sin[2(k-r+1)\theta_{0}\pm2(n+m+1)\zeta],  \nonumber \\
	\Omega_{J}&=&\pm [(n-m)/(k-r)]\Omega  \Longrightarrow I^{9}_{s}=\frac{1 }{\phi_{s} R} \varSigma_{3}\sin[2(k-r)\theta_{0}\pm2(n-m)\zeta],  \nonumber \\
	\Omega_{J}&=&\pm [(n+m+1)/(k-r)]\Omega \Longrightarrow I^{10}_{s}=\frac{1 }{\phi_{s} R} \varSigma_{3}\sin[2(k-r)\theta_{0}\pm2(n+m+1)\zeta],
	\label{eq:IS_genera}
	\end{eqnarray}
	where
	\begin{eqnarray}
	\varSigma_{1}&=& \sum_{m=0}^{\infty} \sum_{n=0}^{\infty} \sum_{p=0}^{\infty} \sum_{k=0}^{\infty} \frac{(-1) ^{p}(2n+2p)_{2k}  \xi^{2k}  (\phi_{s} R)^{2p+2n+1}}{2^{2p+2n+1} \Gamma(p+2n+2) (2n+1)!2^{2k+1}(k!)^{2}},
	\nonumber \\
	\varSigma_{2}&=&\sum_{m=0}^{\infty} \sum_{n=0}^{\infty} \sum_{p=0}^{\infty}\sum_{k=0}^{\infty}  \sum_{r=0}^{k-1}\frac{(-1) ^{p} (2n+2p)_{2k}\xi^{2k} (\phi_{s} R)^{2p+2n+1}}{2^{2p+2n+2k+1} \Gamma(p+2n+2) (2n+1)!(2k-r)! r!},\nonumber \\
	\varSigma_{3}&=& \sum_{m=0}^{\infty} \sum_{n=0}^{\infty} \sum_{p=0}^{\infty}\sum_{k=0}^{\infty}  \sum_{r=0}^{k} \frac{(-1) ^{p}(2n+2p)_{2k+1} \xi^{2k+1} (\phi_{s} R)^{2p+2n+1}}{2^{2p+2n+2} \Gamma(p+2n+2) (2n+1)!4^{k} (2k-r+1)! r!}.
	\end{eqnarray}
	The results from Eq.(\ref{eq:IS_genera}) lead to even, odd and fractional current steps in the IV characteristics, depending on the integers values of $m,n,r$ and $k$.
\end{widetext}

\end{document}